# Clickbait in YouTube: Prevention, Detection and Analysis of the Bait using Ensemble Learning

Peya Mowar*

Department Information Technology, Delhi Technological University, peyajm29@gmail.com

Mini Jain*

Department Information Technology, Delhi Technological University, mini.j2799@gmail.com

Ruchika Goel

Department Information Technology, Delhi Technological University, ruchika2ar@gmail.com

Dinesh Kumar Vishwakarma

Department Information Technology, Delhi Technological University, dinesh@dtu.ac.in

Unscrupulous content creators on YouTube employ deceptive techniques such as spam and clickbait to reach a broad audience and trick users into clicking on their videos to increase their advertisement revenue. Clickbait detection on YouTube requires an in-depth examination and analysis of the intricate relationship between the video content and video descriptors – title and thumbnail. However, the current solutions are mostly centred around the study of video descriptors and other metadata such as likes, tags, comments, etc., and fail to utilize the video content, i.e., both video and audio. Therefore, we introduce a novel model to detect clickbaits on YouTube that consider the relationship between video content and title/thumbnail. The proposed model consists of a stacking classifier framework composed of six base models (K-Nearest Neighbours, Support Vector Machine, XGBoost, Naive Bayes, Logistic Regression, and Multilayer Perceptron) and a meta-classifier (Random Forest). The developed clickbait detection model achieved a high accuracy of 92.89% for the novel BollyBAIT dataset and 95.38% for Misleading Video Dataset (MVD). Additionally, the stated classifier does not use meta-features or other statistics dependent on user interaction with the video (e.g., the number of likes, followers, or comments) for classification, and thus, can be used to detect potential clickbait videos before they are uploaded, thereby preventing the nuisance of clickbaits altogether and improving the users' streaming experience.

## 1 INTRODUCTION

The advent and exponential rise of social media, like Twitter, Instagram and YouTube provided their users the opportunity to both create and consume a variety of online content to stay up-to-date with current world events, share their views/knowledge, market their business products etc. Unfortunately, unscrupulous content creators misuse these platforms to publish misleading, poor-quality content which often disseminates false

* Peya Mowar and Mini Jain have contributed to this work equally as first authors.

information[11][21] by employing techniques such as hoaxes[42][46] rumours[24][23][34][45], fake news [25][15][32][43] and clickbaits[42][7]. While the motivations for publishing hoaxes and rumours range from scaremongering to manipulation of popular public opinion, the primary incentive behind posting clickbait content is to gain more clicks on either the content itself (e.g., video) or the link attached to the content (e.g., news headline with link to full article). An increased number of clicks leads to an increase in advertisement revenue for the creator but degrades the quality of user experience and often leaves users feeling unsatisfied, tricked, and annoyed. In the context of YouTube, examples of such content include videos with morphed eye-grabbing thumbnails, exaggerated headlines that include curiosity-inciting phrases such as “You Won’t Believe”, “X Reasons Why” and often fail to deliver on the promised sensational content, or the videos whose content is completely irrelevant or significantly digresses from that suggested by the video descriptors.

Clickbait as a concept was originally found in news media, where captivating headlines had a significant dissonance from the actual content of the article [5][6] .With the increasing popularity of social media, these text-centric baits were additionally seen in platforms such as Twitter and Reddit. As a consequence, researchers were prompted to develop clickbait detection models for identifying clickbaits in text-centric social media [1][44][28]. Subsequently, there has been a shift from text-centric baits, to visual-centric and video-centric baits, wherein content creators make use of enticing thumbnails and videos to grab a user’s attention respectively. This motivated researchers to redefine clickbaits to encompass their different types and develop models to detect them on various other types of social media such as Instagram [12] and YouTube [41][31][38]. Previously, we explored visual-centric baits and created a cross-platform clickbait detection model comprising of a stacking framework architecture [17]. Encouraged by the outcome, we decided to foray into detection of video-centric clickbaits.

Since, clickbait detection and prevention in videos is a largely unexplored field, the primary aim of the conducted research is to develop the Clickbait Prevention and Detection Model (CPDM) for YouTube videos based on only the (i) Video Content, i.e., both the video and associated audio, (ii) Video Title, and (iii) Thumbnail. The video title and thumbnail are used by the viewer to decide to click on a video and, ideally, should be an accurate representation of the video content. The model will determine if they have been exaggerated, manipulated, and engineered to entice the users into clicking on them or are ‘clickbaity.’ Besides, we aim to utilize the information and features extracted from the descriptors and the video to build a model that can determine if the latter fulfils the promise of content suggested by the former. This contrasts with the previously developed models for clickbait detection on YouTube which are heavily dependent on meta-features such as comments, number of likes etc. for the video accumulated over a considerable period of time.

Therefore, to the best of our knowledge, CPDM will be the first model that not only takes into account the intricate relationship between the video contents and its title/thumbnail utilizes it to classify a video as clickbait, but is also completely independent of any meta-features or other features dependent on user engagement with the video. Hence, it can be used for real-time detection of clickbait videos and be used to discard them before they reach a broad audience and lead to poor user experience. Hence, while previous research [41][31][38] focused on developing clickbait detection models, i.e., ‘detect’ them after they have been published and garnered significant user engagement, the proposed model can be used to both ‘prevent’ the publishing of clickbait content and deterioration of content quality and ‘detect’ it if it has already been published.

In our research, we have defined the videos on YouTube as clickbait if they fall into one or more of the following five categories:

1. Misleading content: These contain false information or fake news in either the video descriptors, i.e., headline and thumbnail and/or video content, e.g., A video showing morphed images of an actress's baby bump when she is not pregnant.
2. Spam Content: These videos use tricks such as referring to many famous figures in title and thumbnail that are irrelevant to the content to mislead them into clicking on videos to increase ad revenue or redirect them to malicious websites, e.g., A video with title 'Sultan Full Movie | Salman Khan | Aamir Khan | Katrina Kaif' but the video redirects to a link to watch the full movie instead.
3. False Promises: These videos fail to deliver on the content indicated by the headline or thumbnail, e.g., A video that claims to be the official trailer of an upcoming movie but instead shows press meetings of the lead actors.
4. Exaggerated Headline/Thumbnail: These videos contain exaggerated descriptors that are often embellished with emojis, capitalized words, sentimental and catchy phrases, arrows to manipulate the user into clicking on them, e.g., A video with headline, '5 Extremely Bizarre Indian Movie Scenes That Will Surely Leave You SPEECHLESS | Watch Alone'.
5. Exploits Curiosity Gap: These videos contain descriptors that reveal just enough information to incite the user's curiosity but not enough to satisfy it unless they click on the video, e.g., A video with the headline 'This World-Famous Hollywood Actor got Secretly Married Yesterday | Find out WHY and WHERE' and with a video thumbnail that contains a person's blurred image to arouse the user's interest further.

Examples of YouTube videos for each category are shown in Figure 1.

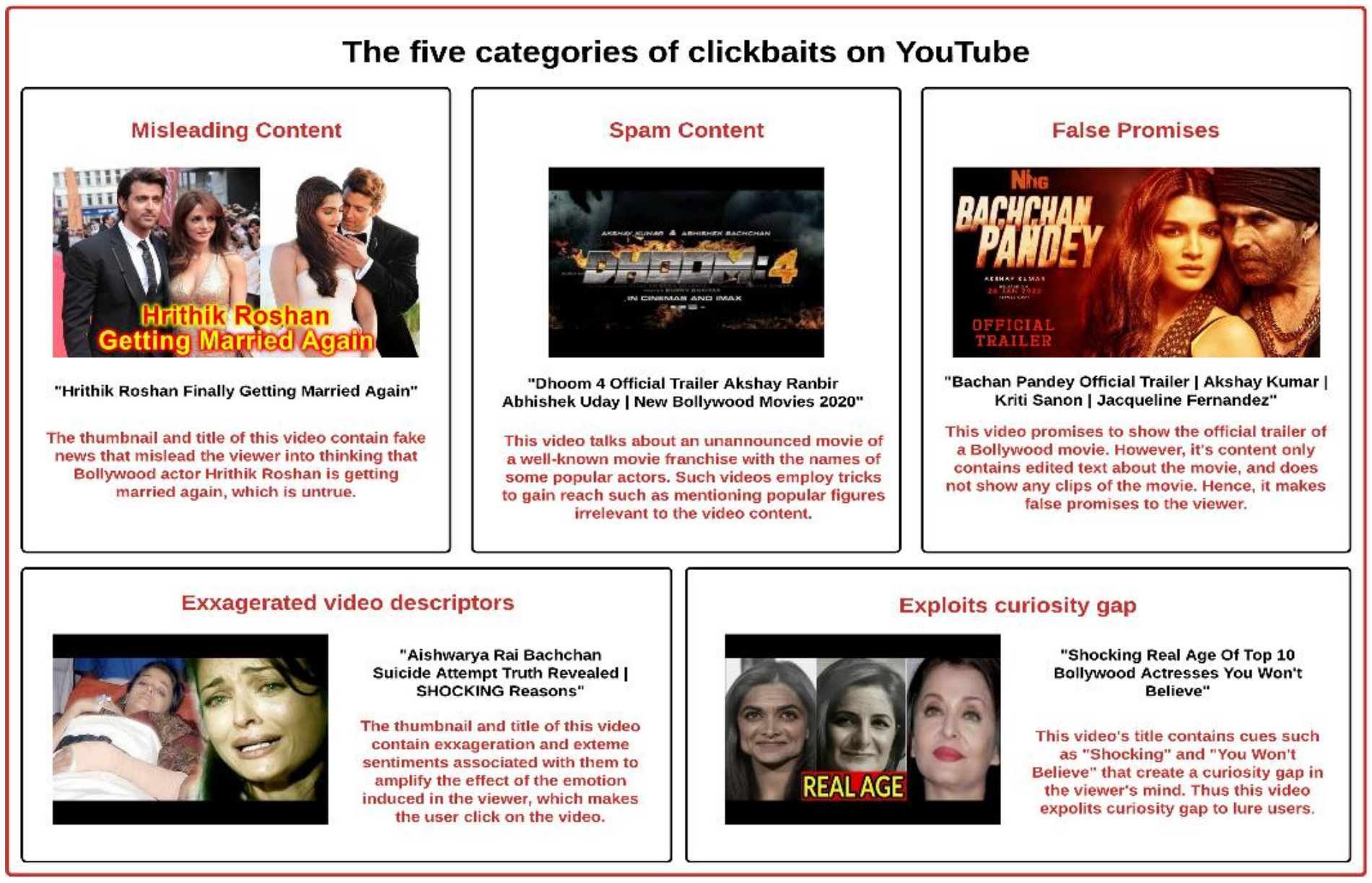

Figure 1: The five categories of clickbaits on YouTube.

The major contributions of our research include:

- Clickbait ‘Prevention’ and ‘Detection’: CPDM uses only the video’s keyframes, its audio, and associated title and thumbnail to classify a video as clickbait or not and is independent of any meta-features or user-dependent features of the video. Since, all features used by CPDM are available at the time of video upload, our model can not only detect existing clickbaits on YouTube but can also be used to identify a clickbait video in real-time before it is made available to public, and thus prevent the problem of clickbait altogether.
- Multilingual Support: CPDM supports identification of clickbait videos in over 200 languages (e.g., English, Hindi, Telugu etc.) and thus, has a larger reach than previously proposed models which can mostly detect clickbait videos in English.
- BollyBAIT Dataset: A novel manually annotated dataset that contains 1000 videos belonging to clickbait and non-clickbait categories. Clickbait videos have been further classified into one or more of the types of clickbait defined above. The videos have been selected from matters related to Bollywood because it is India’s leading entertainment industry with a worldwide audience and reach [47] and hence, draws many clickbait creators and spammers. Also, it is multilingual and contains videos in English, Hindi, or a mixture of both.
- Redefinition of Clickbaits on YouTube: Since, social media platforms such as YouTube are ever-evolving, in terms of technology, available content, its creators and consumers, the accompanying forms of clickbaits are also rapidly expanding. Consequently, even though researchers in the past have attempted to define clickbaits [41][38], these definitions are outdated and not broad enough to encapsulate all types of clickbaits currently present on YouTube. Therefore, we performed a holistic study of clickbaits on YouTube and its previous definitions, to identify its various new and emerging categories and then redefined clickbaits to include a vast majority of them. The updated and broader definition of clickbaits was ultimately incorporated in the proposed BollyBAIT dataset and model, which to the best of our knowledge, can now detect and help in preventing a wide majority of currently prevalent and emerging clickbait techniques.

This paper is organized as follows. Section 2 reviews the related work on defining clickbait for YouTube, various approaches to clickbait detection, research gaps, and future scope. Section 3 introduces the novel clickbait dataset, BollyBAIT, and discusses its features and applications. Section 4 discusses the methodology for feature extraction from video and its descriptors and presents an overview of the architecture of our proposed model, CPDM. Section 5 discusses the results of clickbait identification on BollyBAIT and MVD Dataset using the proposed model. Finally, Section 6 presents the conclusion and explains the future scope for the conducted research.

## 2 LITERATURE REVIEW

Much of the previous work in clickbait YouTube detection has focused on detecting text-centric clickbaits in articles. However, with the emergence of visual-centric social media and video-streaming platforms like YouTube and Vimeo, newer forms of clickbaits in terms of images and videos surfaced. As a result, researchers redefined clickbaits and proposed models to cater to these newer modalities. This section presents an overview of past definitions of clickbaits and developed schemes to detect them on YouTube.

## 2.1 Defining Clickbait for YouTube

Due to the ever-evolving nature of social media, the definition of clickbait has evolved too as the intended user-base, use-cases, and the supported content modalities of these platforms diversified. While the early interpretations of clickbaits focused on tabloid journalism, where clickbaits referred to short messages that enticed users to click on the attached link [18], this description was narrow to encompass the variety of clickbaits that came into existence as visual-centric social media became popular. The posts on these contained more descriptors than short textual messages in images, teaser videos, and audios. Therefore, the need for a broader definition of clickbaits that took into consideration the new modalities arose.

In the case of Instagram, clickbaits have been defined as incongruous pairs of text and accompanying images that have been purposely architected to lure users into viewing advertisements and other content to increase their engagement and revenue [12]. For YouTube, researchers have employed various definitions of clickbaits. Zannetteou et al. [41] have defined clickbaits as videos that fall into one or more of the following categories: (i) using dramatic, enticing, and content-irrelevant thumbnails (ii) embedding fake news or false information into the title, thumbnail, or video, (iii) using exaggerated headlines to incite user curiosity and lure them into clicking. However, Shang et al. [31] have defined clickbaits as videos whose content sharply differs from the one promised by the title or thumbnail. Varshney et al. [38] have based their categorization of clickbait on the one proposed by George Loewenstein and defined them as videos that exploit the information gap theory to entice a user into clicking on a video that does not contain the suggested content and eventually degrades the user experience.

## 2.2 Detecting Clickbaits on YouTube

There have been primarily two approaches to clickbait detection in YouTube, one that took into consideration the content of the video and the other that relied only on video descriptors like title, thumbnail, and other metadata like comments, the number of views, and likes to detect clickbaits. Below, we discuss the past work done in terms of both these approaches.

### *2.2.1 Content-Agnostic Approaches for Clickbait Detection*

Zannetteou et al. [41] developed a deep-learning-based model based on variational autoencoders that do not directly analyse the video content but uses features extracted from data and video descriptors such as headline, tags, thumbnail, comments, number of views, likes, etc. which were collected using the YouTube's Data API for 206k videos. The extracted features were majorly text-based or statistics except for thumbnail image. On the other hand, Shang et al. [31] have adopted a fully content-agnostic approach and only used comments and metadata (e.g., number of views) while building the Online Video Clickbait Protector Scheme (OVCP). Besides not using video content, they did not utilize the descriptors available to users before clicking on the video, such as image and thumbnail, unlike [41].

### *2.2.2 Methods Utilising Video Content for Clickbait Detection*

While most research has focused on developing clickbait detection models utilizing video descriptors and meta-features, only a few recent approaches have used features extracted from the video content, i.e., audio and video, to classify clickbaits. Varshney et al.[38] have incorporated the cosine similarity between the transcript of the video's audio and its title as a feature to inspect if the video delivers on the content as suggested by the title. However, researchers have not employed any features based on the visual aspect of the video in training their models previously.

### 2.3 Research Gaps

In [41], Zannetteou et al. have concluded that the YouTube Recommendation System fails to consider clickbaits while recommending related videos and is likely to recommend more clickbait videos when users click on one such video. Therefore, this lacking feature of the YouTube Recommendation System pushes the user into a web of clickbaits, wasting their time and causing user dissatisfaction. Furthermore, they examined two statistics for both clickbait and non-clickbait videos: (i) Number of Unavailable Videos (NUV): Number of videos removed by YouTube or video publisher (ii) Mean of Time Elapsed (MTE): The mean of time elapsed from video publication date till January 10, 2017, for all videos in each category. After performing this analysis, they deduced that YouTube employs no timely corrective measures to remove clickbait videos because of the small percentage of NUV for clickbaits and a high value for MTE.

Accordingly, a model that can prevent clickbaits altogether, i.e., label a video as clickbait before uploading and discoverable by the public, can help tackle the menace of clickbaits on YouTube and improve user experience immensely. This can be achieved by utilizing only the modalities available during upload, like video content (both audio and video), its title, and thumbnail, and not relying on the meta-data generated due to user interaction with video like the number of likes, dislikes, comments, etc. Furthermore, authors in [31] draw attention to their analysis and subsequent conclusion that focus and utilization of only one of the titles, thumbnail [37][40], or the video content [27] cannot be a definitive indicator of the video being a clickbait. Consequently, there is scope to design a model that explores their complex relationship between video content and title/thumbnail.

## 3 BOLLYBAIT: A BOLLYWOOD-CENTRIC CLICKBAIT DATASET FOR YOUTUBE

We present a novel clickbait dataset with 1000 videos for YouTube titled 'BollyBAIT' containing 500 clickbait and 500 real/non-clickbait videos. All videos in the dataset are related to Bollywood and belong to one or more of various categories such as video interviews of celebrities, trailer launch videos, songs, trailers of Bollywood movies, and news and updates related to ongoing happenings in the industry or Bollywood celebrities' lives. The primary reason for focusing on Bollywood-related videos on YouTube is the popularity of the Bollywood industry. The Indian film industry, especially Bollywood, attracts a lot of attention [47] with a worldwide audience [48]. Therefore, many unscrupulous content creators who want to draw mileage from the popularity of Bollywood to gain advertisement revenue create spam and clickbait videos to reach a vast audience, resulting in an enormous number of clickbait videos available on YouTube.

The videos have been manually collected by searching for queries such as "Upcoming Bollywood movie trailers", "Anushka Sharma pregnancy", "Full movie Bollywood", "Latest news Shah Rukh", "Shocking news Kareena Kapoor". The queries were formed by including famous Bollywood celebrities like Anushka Sharma, Kareena Kapoor, Shah Rukh Khan, and phrases of interest like pregnancy, latest movies, shocking news, etc. After collecting the videos, they were manually labelled by three reviewers as clickbait or real by analysing the (i) Title, (ii) Thumbnail, and (iii) Video content. For each video, it was examined if they belonged to the categories that were defined earlier in Section I of the paper. A video was labelled as clickbait if it fell into one or more of these categories; otherwise, it was labelled as real. Our dataset covers a massive number of clickbait videos related to Bollywood as shown in Figure 2.

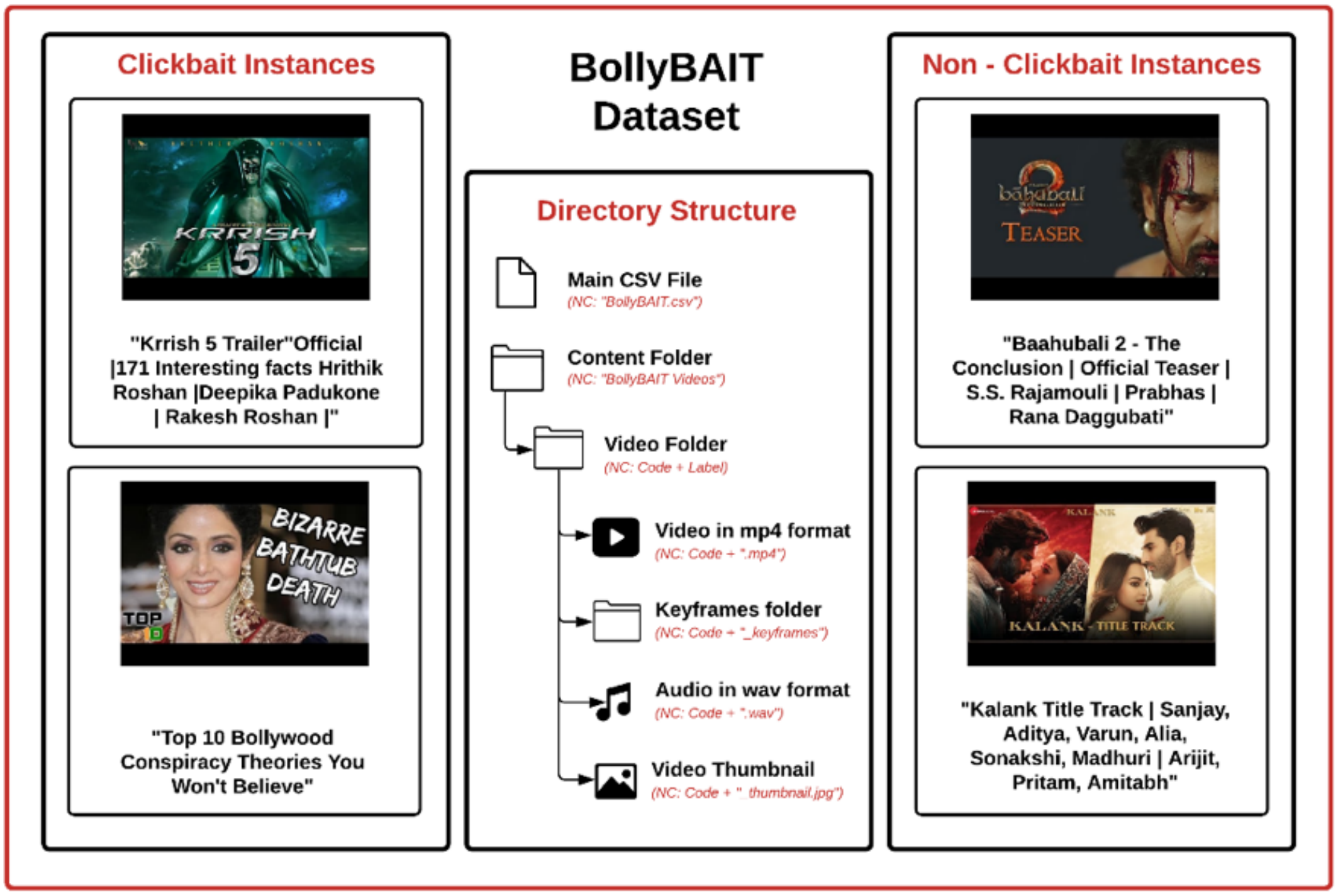

Figure 2: BollyBAIT dataset instances and folder structure

The features of our dataset include: (i) Multilingual: The videos in our dataset belong to many languages such as Hindi, English etc. and therefore provide an opportunity to develop a language-independent model for clickbait detection (ii) Manual Annotation: All the videos have been manually annotated and reviewed by multiple reviewers which allowed us to incorporate categories of clickbait that examine the intricate relationship between the video and its title/thumbnail, currently only possible by human examination such as if the video exploits the curiosity gap of the user and contains false promises or not (iii) Multiple Modalities: In labelling a video as clickbait or not, we have considered four different modalities available: (a) audio of the video, (b) video, (c) text i.e. title of video, and (d) image i.e., thumbnail and extracted features from all of them, often converting one modality into the other (iv) Encompasses a broad definition of clickbait: Our dataset employs a varied and broad definition of clickbait by the virtue of five different categories of clickbaits considered.

## 4 PROPOSED METHODOLOGY

In this section, we discuss the proposed YouTube clickbait detection framework in detail. The pipeline of the proposed framework for CPDM is composed of the following steps as shown in Figure 3: (i) Pre-processing: This involves extraction of video and associated thumbnail image, title, audio, and keyframes (ii) Feature extraction: Features are extracted from the teaser text (i.e., the Title) and image (i.e., the Thumbnail) of the video using the pre-trained Bert-Base and Resnet-50 models respectively (iii) Feature set mining: The 6 different feature sets obtained from different modalities are mined using various deep learning techniques (iv) Fusion: Finally, the results of all the deep learning models are combined to predict whether a video is clickbait or not. Each of the steps in the proposed framework are discussed in detail in the following sub-sections.

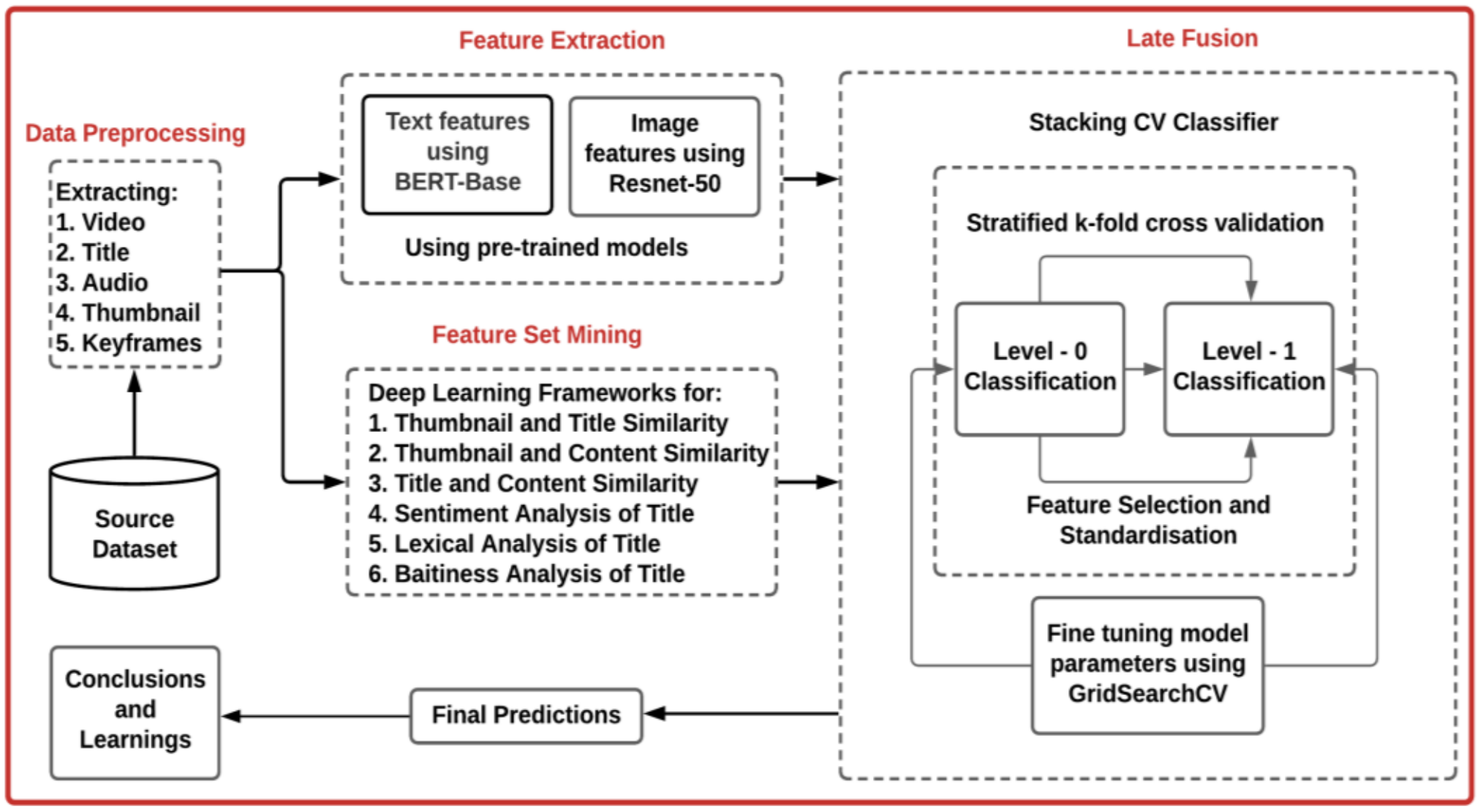

Figure 3: Clickbait Prevention and Detection Framework (CPDM)

---

**ALGORITHM 1: Clickbait Detection and Prevention Framework (CPDM)**

---

for each video in dataset, do
    extract video, audio, thumbnail, title and keyframes
    extract title text features using BERT-Base and thumbnail image features using Resnet-50 models
    perform feature set mining to extract video features, do
        generate a graph representation for the video to determine title-content disparity
        generate image caption for thumbnail and compare with title
        transcribe the audio content and compare with title
        perform lexical, baitiness and sentiment analysis on the video title
    train and test a stacking classifier on these features with manually annotated label for the video
    return predicted classification label for the video: clickbait or non-clickbait
end

---

### 4.1 Data Pre-processing

Video components and descriptors of various modalities i.e., thumbnail, video, audio and title text in the source dataset were scraped using the Python YouTube Data API Library [49]. Subsequently, video keyframes which are the descriptive and essential frames of a video were extracted using the FFMPEG open-source program [50][35] and the video's audio was obtained as a .wav file using the YouTube-dl library [51]. Finally, a wide variety of features representing the various modalities present in a video were extracted using these various components and fed to the proposed model for training.

### 4.2 Feature Extraction

Text feature extraction from the teaser text, i.e., the title, was performed using the BERT-Base (Bidirectional Encoder Representations from Transformers-Base) pre-trained model [8]. It is devised to pre-train deep bidirectional representations from an unsupervised text by simultaneously conditioning on both left and right context. Consequently, the BERT Base model can be fine-tuned by adding merely one supplementary output layer to create classifiers for a broad array of NLP assignments such as Sentiment Analysis [14] and Next Sentence Prediction Tasks [33]. In addition, the BERT-Base model can extract conceptualized word embeddings for the text from the title to get the pooled output of 768 units as the feature vector. Image features from the thumbnail associated with the video were extracted using ResNet50 [13] which is a 50-layers deep convolutional neural network trained on more than a million images from the ImageNet database [10], and can classify photos into a thousand target categories. Consequently, the network has acquired deep characteristic representations for an extensive collection of images and, in most cases, can be used without any fine-tuning. Accordingly, it was used to extract image features from the image to get the pooled 2048 units as the feature vector.

### 4.3 Feature Set Mining

Six prominent feature sets were mined based on thumbnail, title, and video content using various deep learning techniques as follows:

- Content-Thumbnail Disparity: It uses a graph-based deep neural network to produce a vectorized representation of the disparity between the video's thumbnail and the actual content, i.e., the frames of the video.
- Thumbnail-Title Disparity: A text caption for the thumbnail generated using an image captioning neural network was compared with the video's title to get a representation of the disparity between the two texts.
- Audio-Title Disparity: It uses an elaborate model to process audio, transcribe it and then compare the transcribed text to video’s title.
- Sentiment Analysis of Title: The sentiment score of the title was calculated based on its negativity/positivity/neutrality and intensity.
- Lexical Analysis of Title: It is a count-based feature quantifying the presence of lexical cues like capitalized letters or words, excessive punctuations etc. to grab user’s attention.
- Clickbaitiness of Title: It is a count-based feature used to quantify the presence of various enticing and curiosity-generating words and phrases in title which ultimately lure the user to click on the video.

#### *4.3.1 Content-Thumbnail Disparity Module*

To analyse the similarity between the video's thumbnail and content, we constructed a framework based on the classification of videos represented as graphs labelled clickbait or non-clickbait. Each video has a weighted directed graph associated with it, which comprises of a root node - the thumbnail, and the children nodes – the keyframes. Each edge from the root to a child captures the resemblance between the two nodes, hence the graph captures the correlation between the thumbnail and video content. A sample graph has been shown in Figure 4 wherein the thumbnail image is the root node connected to all the child nodes, i.e., the keyframes.

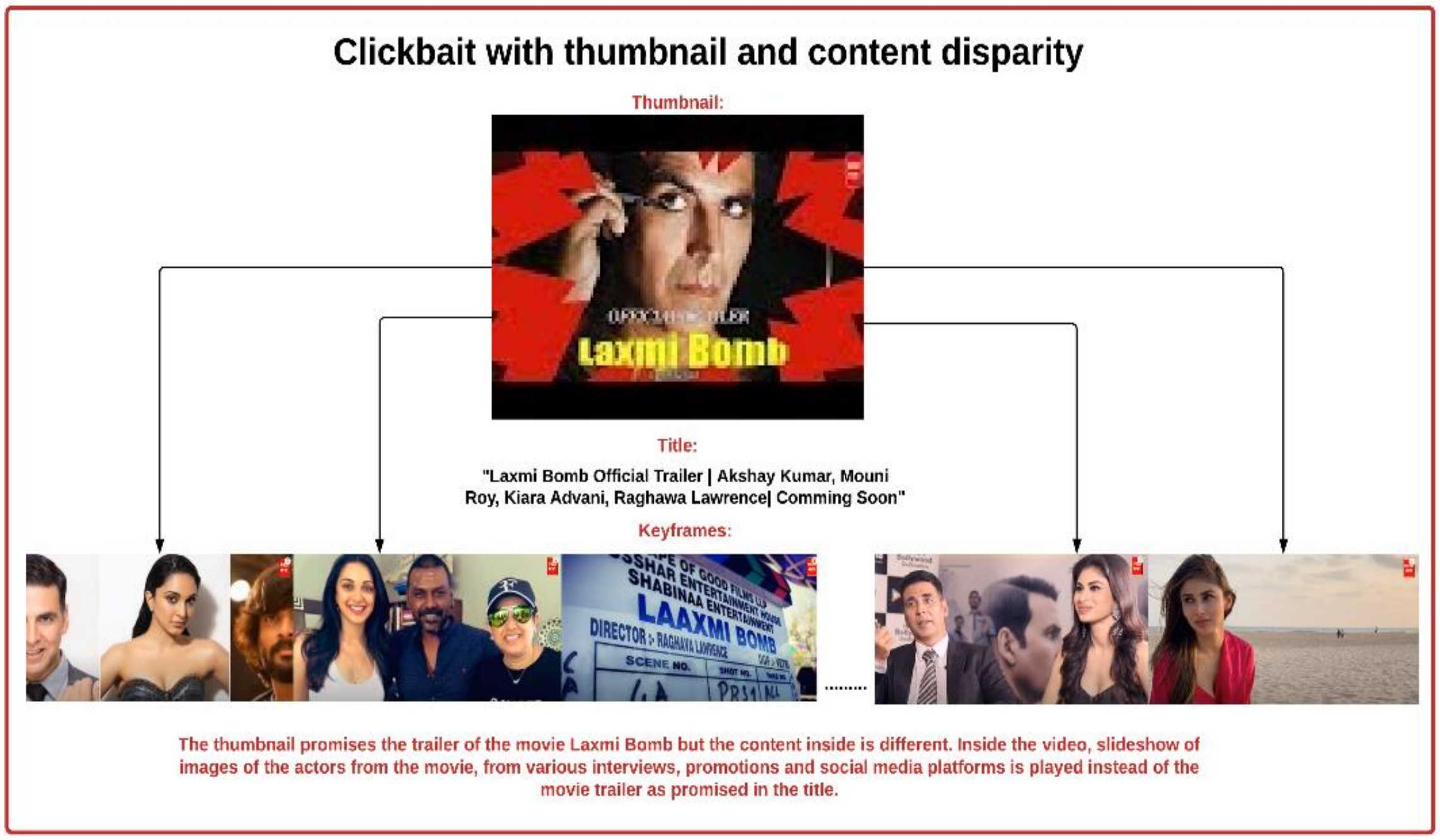

Figure 4: Example of a clickbait video with thumbnail and content disparity

The weight of each edge is calculated as the summation of the following three measures of similarity between the thumbnail and that keyframe: (1) Facial recognition, (2) Object recognition, and (3) Visual image similarity, which have been explained as follows:

1. Facial recognition: It is an empirical expression representing the intersection of faces detected in both keyframe and the thumbnail. Faces present in the thumbnail ($F_{TH}$) were detected using the DNN Face Detector model in the OpenCV library which is based on the Single-Shot Multibox Detector and the facial embeddings for the same were stored [22]. Then, while processing each of the keyframes, all the faces detected in the keyframe ($F_{KF}$) using the same MultiBox Detector technique were matched against all those in the thumbnail using the stored facial embeddings for each of the thumbnail, to obtain the number of common faces (CF). This matching is computed by evaluating the Euclidean distance between the two candidate embeddings. Subsequently, the weight was obtained by calculating the relative number of matched faces against the total number of faces present in the keyframe and the thumbnail. The mathematical definition of the weight ($W_1$) based on this measure is defined in Eq. 1. It is considered to be undefined, when there are no faces found in both the thumbnail and keyframe, since in such a situation, this weight holds no meaning.

$$W_1 = \begin{cases} \frac{2\,CF}{F_{TH} + F_{KF}}, & F_{TH} + F_{KF} > 0 \\ undefined, & otherwise \end{cases} \tag{1}$$

2. Object Recognition: It is an expression representing the convergence of objects present in both the thumbnail and the keyframe. For this purpose, objects were detected using the RetinaNet object detection model [36] and the lists of objects detected in the thumbnail image ($O_{TH}$) and the keyframes ($O_{KF}$) were extracted. Then, thumbnail and keyframe lists were matched to get the count of common objects (CO) in the two lists and the subsequent weight was defined as this number against the total number of objects present in the keyframe and the thumbnail. This weight ($W_2$) has been defined in Eq. 2, and is undefined when no object was present in both the thumbnail and the keyframe.

$$W_2 = \begin{cases} \frac{2\,CO}{O_{TH} + O_{KF}}, & O_{TH} + O_{KF} > 0 \\ undefined, & otherwise \end{cases} \tag{2}$$

3. Visual Image Similarity: This similarity is obtained by comparing the low-level image features that are dependent on the pixels of the image. For this purpose, colour histogram vectors for both the images were constructed by quantifying the colours within the images and enumerating the number of pixels of each colour. Then from these colour histograms, vectors of 3 units – containing the summation, mean and standard deviation of the bins were obtained. The Manhattan distance (d) was used to calculate the distance between these two image vectors which was used as the measure for visual histogram-based similarity. For further elaboration and quantification of this distance, please refer [30]. Finally, we normalise the value of d to obtain this weight ($W_3$), such that the weight lies in the range of [0,1].

Finally, the mean of these three weights ($W_1$, $W_2$ and $W_3$) is taken to be the edge weight ($\bar{W}$) between the thumbnail and the keyframe, as shown in Eq. 3. Each of these weights lie in the range [0,1] and therefore their mean also lies in the same range. In the situation that either or both weights $W_1$ or $W_2$ are undefined, the mean is taken of the remaining defined weights, and the undefined weights are ignored, as they hold no significance.

$$\bar{W} = \frac{W_1 + W_2 + W_3}{3} \tag{3}$$

The video thumbnail and keyframes were processed and corresponding edge weights were calculated and assigned to create a representative graph for each video. Each node in the graph is also assigned its own set of features which is the feature vector extracted using the Resnet-50 pre-trained model of the nodal image. The algorithm used for the supervised classification of these graphs is the Deep Graph Convolutional Neural Network Algorithm [3]. Graph CNN is commonly used in processing of videos as they vary in size and contain complex temporal and spatial interdependencies [20]. The architecture of the graph classification neural network is shown in Figure 5 which is used to extract a pooled output of 16 units from the last layer.

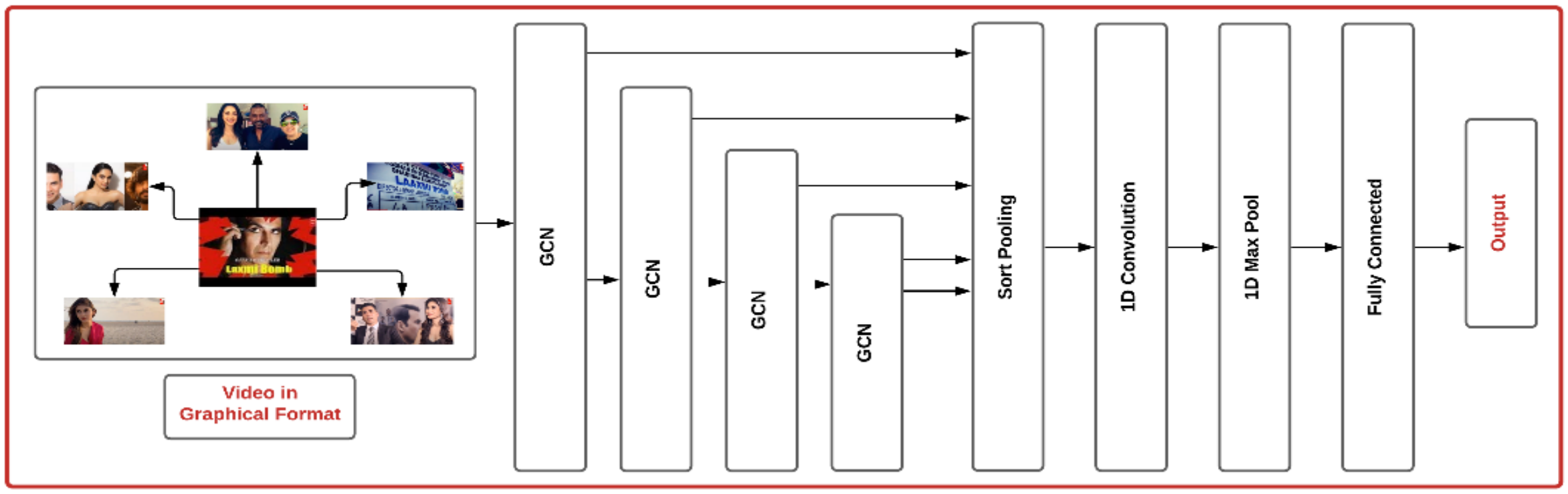

Figure 5: Deep Neural Network based on Graph CNN for identifying thumbnail and content disparity

#### *4.3.2 Title-Thumbnail Disparity Module*

Image captioning has been used to capture the textual description of the thumbnail and compare it with the title. Figure 6 shows a clickbait video with the title "new" and a thumbnail depicting a certificate of a movie that has been released. The title doesn't describe what the video contains, neither does it align with the thumbnail itself. This creates a curiosity gap and misleads the viewer into assuming it is a trailer for a new movie.

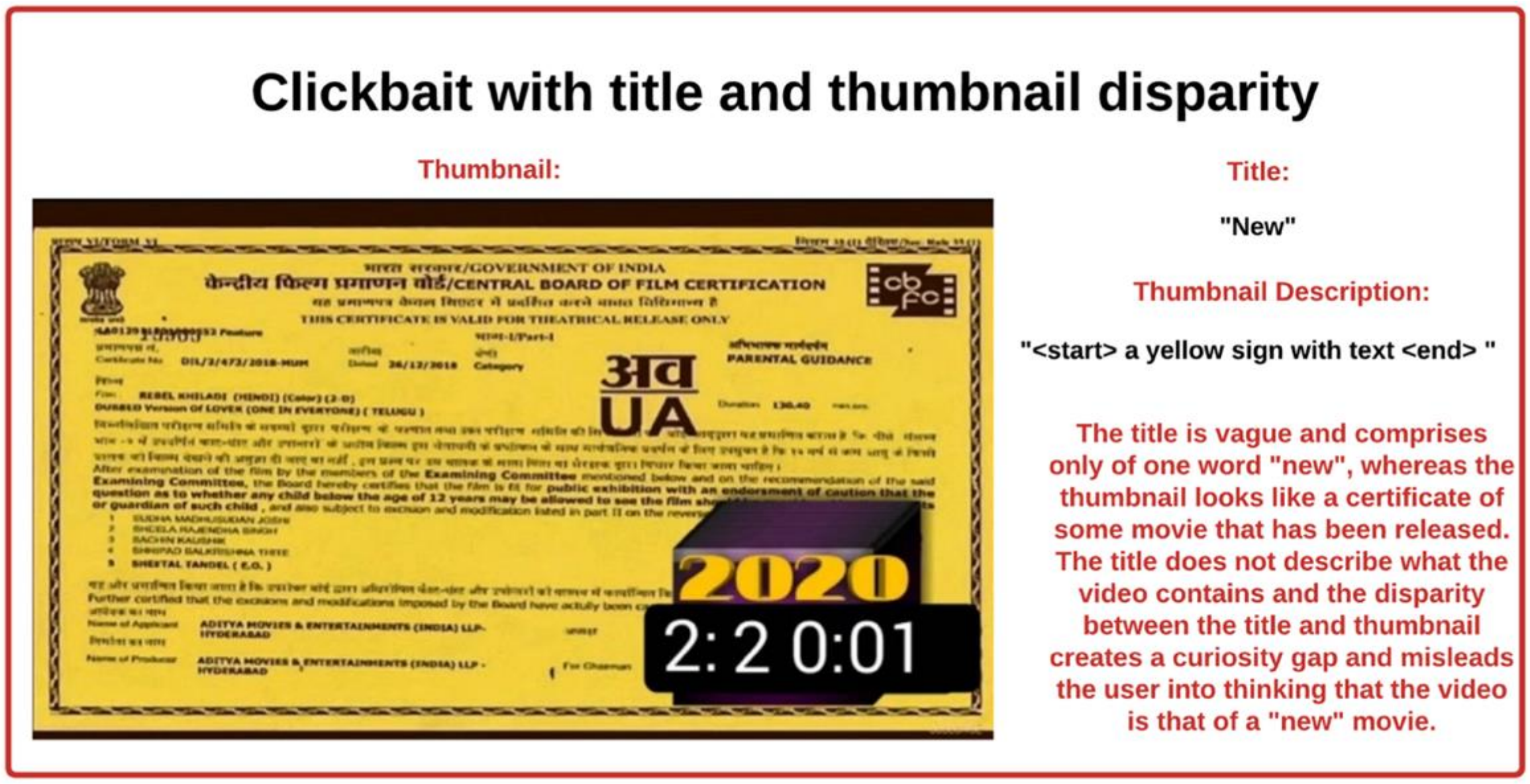

Figure 6: Example of a clickbait video with title and thumbnail disparity

The MS COCO (i.e., Microsoft Common Objects in Context) dataset [19] for image captioning was used to train a neural network-based model for describing for the video thumbnails. It is a large-scale dataset consisting of >300K images with five captions per image. The model contains a CNN encoder which is given images for the task, and its output is fed into the RNN decoder, which gives the generated caption for the image. After getting the predicted caption for a thumbnail, it was compared with the video title that had been translated to English using the Google Trans Python library [52]. Figure 7 shows the architecture of the model used to caption the images. Afterwards, the comparison between the generated caption and video title was made by employing Cosine Matching [53] on three different measures: (i) on plain text, (ii) on the text pre-processed using NLTK library [4] (iii) on GLOVE [54] embeddings. A feature vector of three units is obtained from this component.

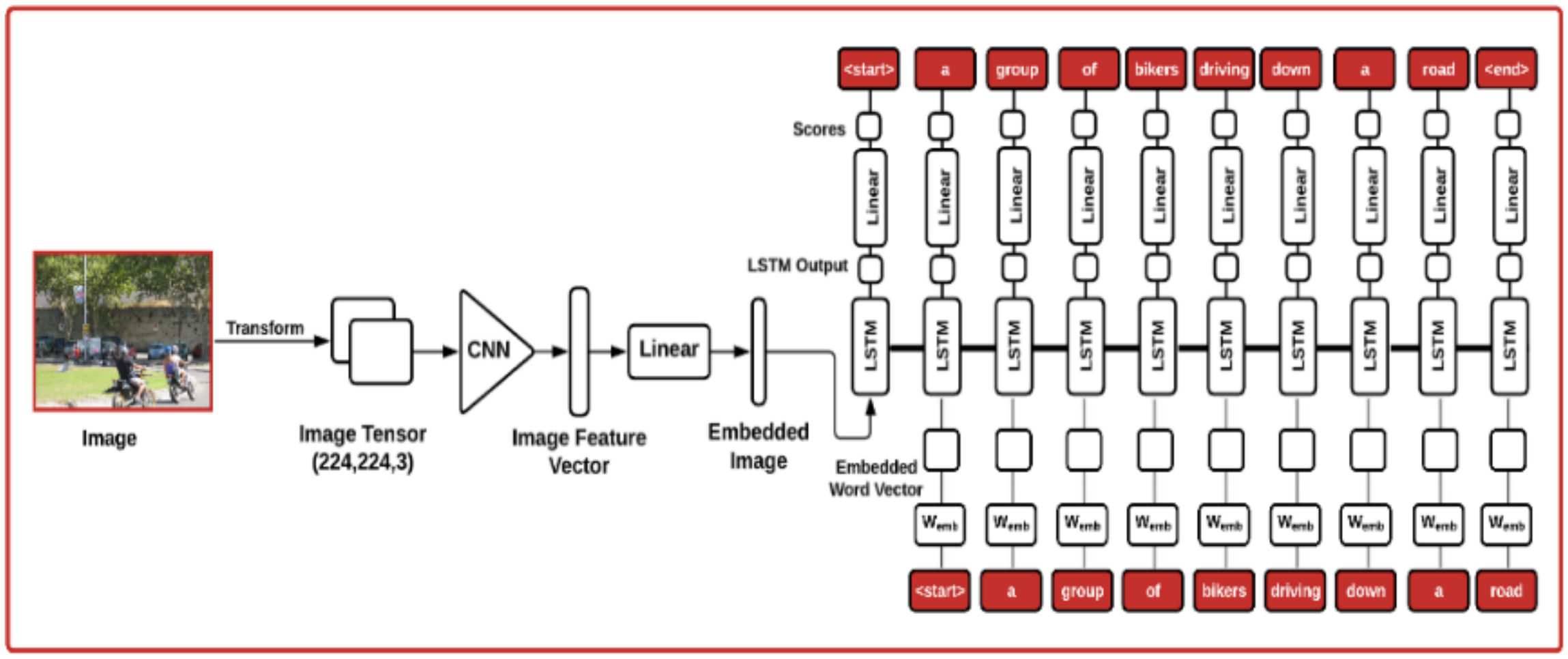

Figure 7: Layers of the deep neural network used for captioning of thumbnail images

#### *4.3.3 Title-Content Disparity Module*

This feature component is based on the correlation between the title, i.e., the promised content and the genuine content of the video. Figure 8 shows an example of a clickbait video wherein the title suggests that Ankita Lokhande adopted a baby named Sushant. However, nothing of this description is mentioned in the audio that was transcribed. This transcription is compared with the title to evaluate the similarity between the title and the video content. The audio was converted to text using the Speech to Text Google Speech API [55]. The transcribed text was then translated to English using the Google Trans Python library as the dataset consists of multilingual videos. Consequently, the comparison between the transcribed audio text and the title was made by employing Cosine Matching, as employed in Section 4.3.2.

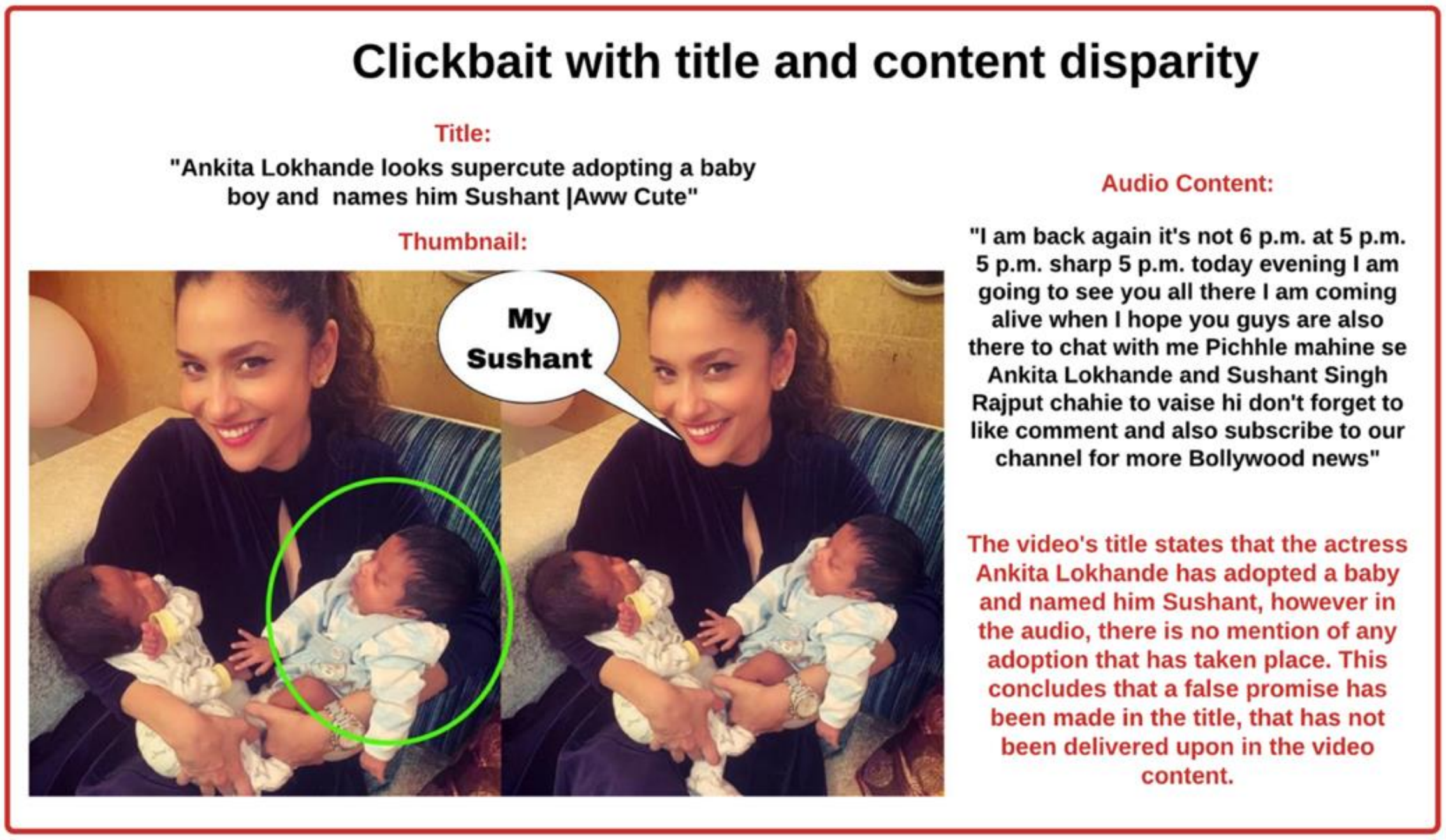

Figure 8: Example of a clickbait video with title and content similarity

### *4.3.4 Sentiment Analysis of the Title*

A helpful attribute that can give important information about the video is the sentiment of its title [29]. Figure 9 shows examples of clickbait video titles using words indicating strong feelings to entice the user into clicking on them. A strong sense of emotion is often used to draw the viewer to click on a video by exploiting the emotional curiosity gap [39]. We have extracted four sentiment scores for each video: positive valence, negative valence, neutral valence, and compound score. A positive valence score ($V_1$) is a measure of the positive sentiment in the title; similarly, negative ($V_2$) and neutral scores ($V_3$) measure those respective sentiments in the title and each of their values lie in the range [0,1]. These scores are computed using heuristics such as punctuation, capitalization, presence of intensifiers, polarity shifting, and polarity negation. A compound score (C) is a measure of the intensity of sensation in the title which is obtained by summing the valence score of each word and applying normalization so that the value lies in [-1,1]. These scores are quantified by the following equations:

$$\sum_{i=1}^{3} V_i = 1 \quad (4)$$

$$C = \frac{x}{\sqrt{x^2 + \alpha}} \quad (5)$$

where x: sum of valence scores of constituent words, and α: normalization constant

To calculate these scores, we have used the VADER (Valence Aware Dictionary for Sentiment Reasoning) [16], a model used for text sentiment analysis that is sensitive to both polarity (positive/negative) and intensity (strength) of emotion. VADER sentimental analysis relies on a dictionary that maps lexical features to emotion intensities known as sentiment scores. Using the VADER deep learning model, a feature vector of 4 units was derived.

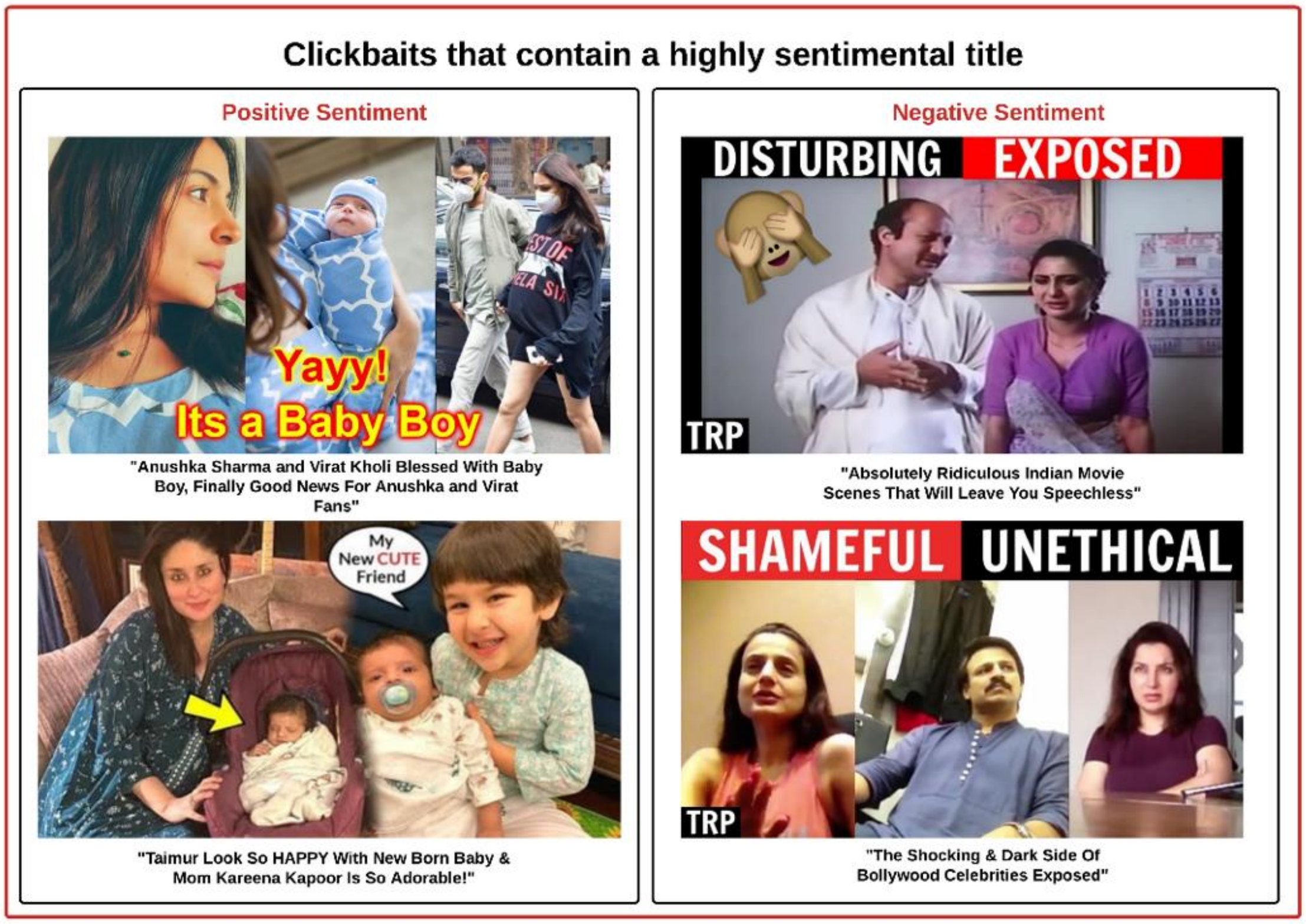

Figure 9: Example of clickbait videos exploiting intensity of a particular sentiment

#### *4.3.5 Lexical Analysis of the Title*

Clickbait video titles use provoking and attention-grabbing cues [26] to entice the user into clicking on them. The most common way to do this is by applying linguistic clues such as overuse of question marks or exclamation marks. Another common occurrence is the use of capital letters to emphasize the enticing words present in the title. Emojis are used, often in excess, to indicate shock or something funny or emotional. Numbers are used for highlighting a good amount of meaty content inside. Figure 10 shows examples of clickbait titles that use these cues to increase the bait value of their video. We extracted the lexical feature vector representing the number of emojis present in the title, presence of numbers in the title, presence of capitalized words in the title, and count of exclamation marks and question marks. These are listed in Table 1.

Table 1: Lexical Features for analysing the title

| Feature | Description |
|---|---|
| If a number present num(x) | Boolean feature which returns true if a number is present in the title and false otherwise. |
| If a question q(x) | Boolean feature which returns true if the title contains a question mark and false otherwise. |
| Count of Emojis e(x) | The count of emojis present in the title. |

| Feature | Description |
| --- | --- |
| Relative Frequency of Capital Letters rf(x) | The ratio of capital letters to total number of letters in the title. |
| Count of Punctuation Marks pun(x) | Returns the count of punctuation marks such as '.','?',',' etc in the title. |

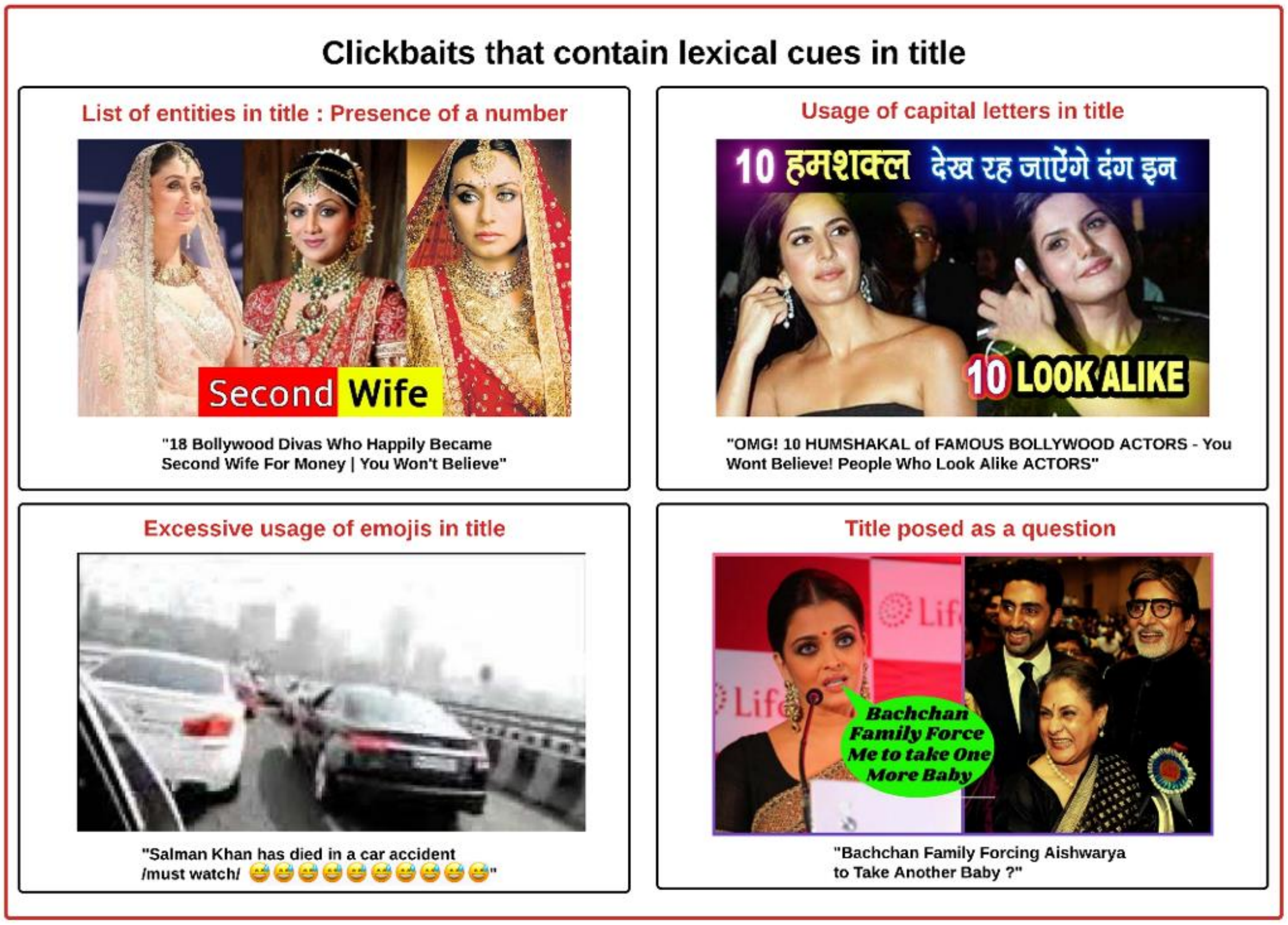

Figure 10: Examples of clickbait videos exploiting lexical cues

#### *4.3.6 Baitiness Analysis of the Title*

Another striking peculiarity common in clickbait videos is enticing phrases or words in the title to lure the user and take advantage of the curiosity gap to increase clicks and views. These include abbreviations like "OMG" i.e., oh my god to express surprise, "LOL" i.e., laughing out loud to describe humor, "ROFL" i.e., rolling on the floor laughing to express laughter. The use of celebrity names or pornographic words [2] like "nudes" or phrases that exploit the curiosity gap such as "you wouldn't believe", "shocking", etc. is also widespread. Figure 11 shows different clickbait video titles that use these various cues to gain views and mislead users. A list of clickbait phrases that we identified to be prevalent in clickbait video titles has been listed in Table 2. A feature vector containing the count of these enticing phrases in the title of the video was extracted.

Table 2: Features for baitiness analysis the title

| Feature | Description |
|---|---|
| Number of celebrity mentions c(x) | This feature represents the count of the number of celebrities mentioned in the title. Celebrities here are the names present in the celebrities' dictionary in the appendix. |
| Count of slangs s(x) | This feature represents the number of slang words present in the title, such as “OMG”, or “WTF”. |
| Count of porn words p(x) | This feature represents the count of porn words present in the title, such as “hot” or “sexy”. |
| Number of Bollywood-focused enticing phrases b(x) | This feature returns the count of alluring phrases specific to the Bollywood industry, such as “Casting couch”, or “Nepotism”. |
| Count of generic luring phrases g(x) | This feature returns the number of captivating phrases present in the title, such as “X Reasons Why” or “You wouldn’t believe”. |

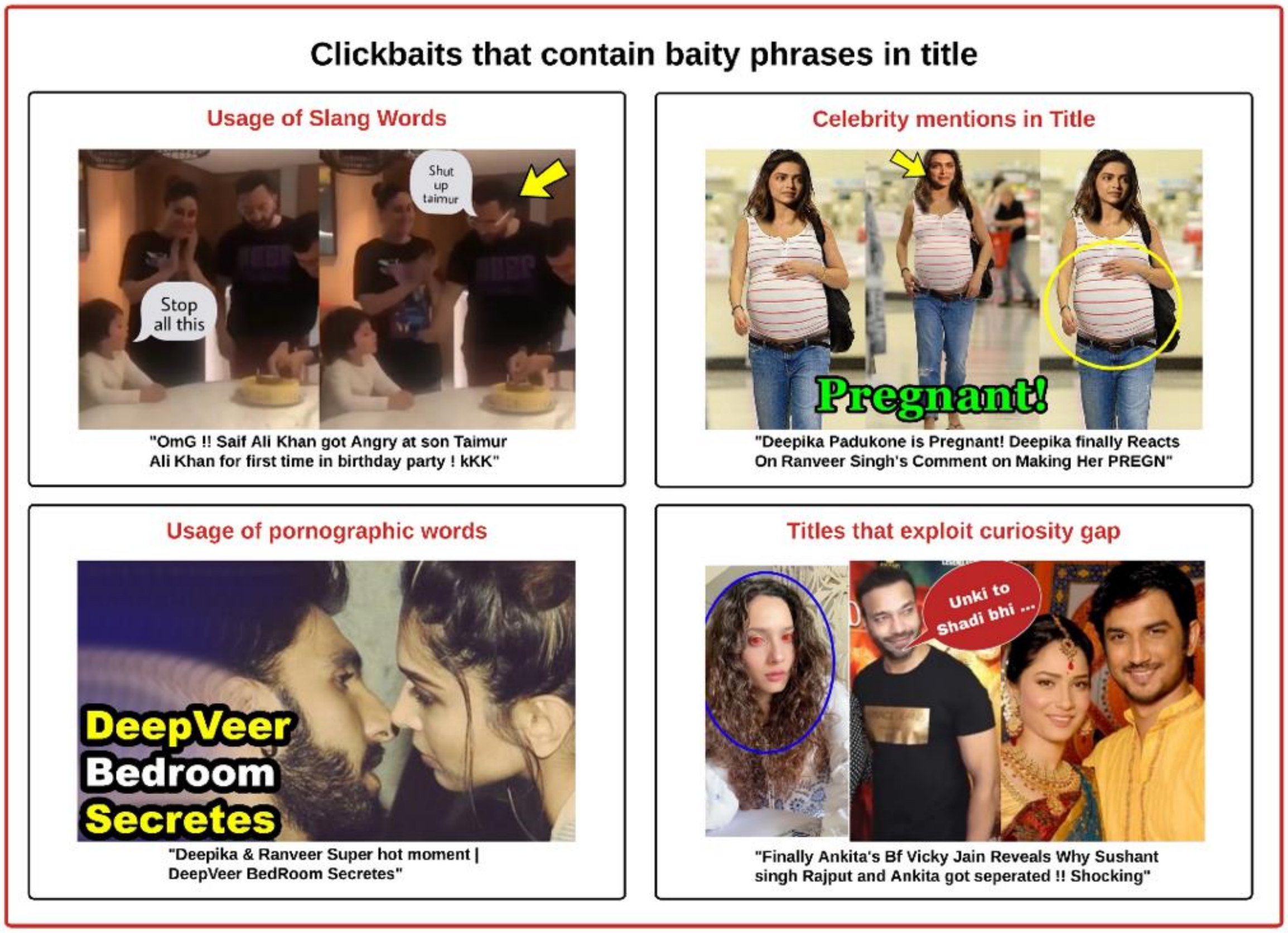

Figure 11: Examples of clickbait videos with entrancing titles to grab attention

### 4.4 Late Fusion

Late fusion is a valuable technique employed while combining prediction scores or feature vectors of different modalities to predict the final output by considering the individual outcomes of all the components or models.

After performing feature set mining and combining all the feature vectors, we performed late fusion by employing a stacking classifier framework consisting of six base classifiers and a meta classifier [9]. Stacking is an ensemble learning method that can be applied for achieving classification via a meta-classifier. The base-level or level-0 classifiers are trained on the complete training set, and then the meta-classifier is trained on the predictions of the level-0 classifier-like features. The base-level model often consists of different machine learning techniques, and therefore stacking ensembles are usually wide-ranging and non-homogeneous. Figure 12 shows the framework of the stacking classifier, with six level-0 classifiers, namely, K-Nearest Neighbours, Gaussian Naive Bayes, Extreme Gradient Boosting Multi-Layer Perceptron, Support Vector Machines and Logistic regression, and the level-1 classifier, Random Forest. We split each dataset into training and test sets (8:2) and carried out 10-fold stratified cross-validation for evaluating the model. We adjusted the parameters to use the top 200 best performing attributes of the training data for classification.

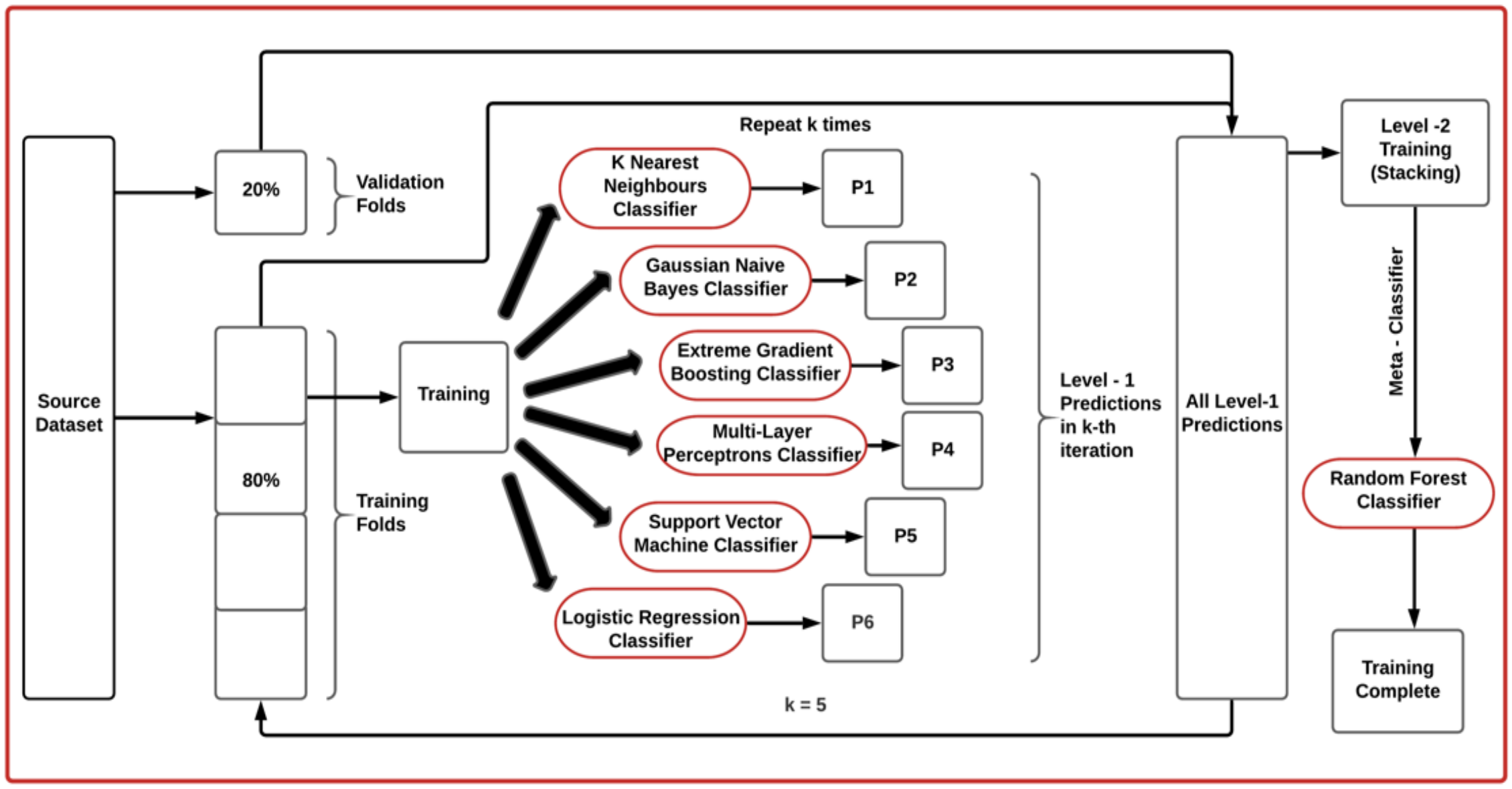

Figure 12: Framework of the stacking classifier

## 5 EXPERIMENTAL RESULTS

In order to evaluate the performance of the proposed algorithm, we conducted experiments on a Bollywood-focused multilingual YouTube dataset (BollyBAIT) and a general YouTube clickbait-focused dataset, which is the MVD (Misleading Video Dataset). We analysed the performance of the CPDM framework on three aspects, namely classification, feature importance and clickbait categorization on each of these datasets, and these results are showcased below.

### 5.1 Classification

This section validates our proposed framework by evaluating it on several datasets and reporting the classification results in terms of several evaluation metrics. For experimentation, we curated a Bollywood-focused multilingual YouTube dataset and combined it with a general YouTube clickbait-focused dataset, which is the MVD (Misleading Video Dataset) [38]. We ran different deep learning algorithms on the two datasets to extract essential features about its telling clickbait characteristics. Subsequently, we trained our stacking classifier on the results from feature mining and evaluated the effects. Table 3 and4 show the fit time (FT), score time (ST), accuracy (ACC), f1-score (F1), precision (PRE), recall (REC), and ROC-AUC (AUC) value for both the BollyBAIT and MVD datasets, respectively, using our stacking classifier and the base models. For both the datasets, the results of the stacking classifier are higher than that of the base models, which proves that the stacking classifier is a better model. All the metrics are the average of the individual metrics over the 10-fold stratified cross-validation run of the classifier. CPDM performed clickbait classification with an accuracy of 92.89% for the BollyBAIT dataset and an accuracy of 95.38% for the MVD dataset. Figure 13 (a) and (b) show the ROC curves for all base models and the Stacking classifier on the BollyBAIT and MVD dataset.

Table 3: Stacking Classifier Results for BollyBAIT

| Classifier | FT | ST | ACC | PRE | REC | F1 | AUC |
|---|---|---|---|---|---|---|---|
| Level – 0: | | | | | | | |
| K-Nearest Neighbours | 30.7 | 0.20 | 0.930 | 0.941 | 0.942 | 0.941 | 0.945 |
| Logistic Regression | 30.9 | 0.20 | 0.924 | 0.933 | 0.940 | 0.936 | 0.975 |
| Gaussian Naïve Bayes | 30.7 | 0.20 | 0.920 | 0.927 | 0.940 | 0.932 | 0.906 |
| Extreme Gradient Boosting | 30.7 | 0.20 | 0.934 | 0.950 | 0.937 | 0.943 | 0.967 |
| Multi-Layer Perceptron | 30.4 | 0.20 | 0.932 | 0.939 | 0.947 | 0.942 | 0.968 |
| Support Vector Machine | 30.5 | 0.20 | 0.929 | 0.941 | 0.940 | 0.940 | 0.976 |
| Level – 1: | | | | | | | |
| Stacking Classifier | 30.4 | 0.20 | 0.928 | 0.941 | 0.932 | 0.936 | 0.976 |

Table 4: Stacking Classifier Results for MVD

| Classifier | FT | ST | ACC | PRE | REC | F1 | AUC |
|---|---|---|---|---|---|---|---|
| Level – 0: | | | | | | | |
| K-Nearest Neighbours | 10.0 | 10.6 | 0.885 | 0.892 | 0.847 | 0.868 | 0.951 |
| Logistic Regression | 86.7 | 00.6 | 0.943 | 0.935 | 0.941 | 0.937 | 0.985 |
| Gaussian Naïve Bayes | 01.2 | 01.9 | 0.791 | 0.766 | 0.771 | 0.766 | 0.888 |
| Extreme Gradient Boosting | 15.7 | 05.6 | 0.877 | 0.890 | 0.829 | 0.858 | 0.956 |
| Multi-Layer Perceptron | 18.4 | 01.1 | 0.957 | 0.947 | 0.959 | 0.953 | 0.987 |
| Support Vector Machine | 80.4 | 13.1 | 0.935 | 0.938 | 0.919 | 0.927 | 0.982 |
| Level – 1: | | | | | | | |
| Stacking Classifier | 25.1 | 12.7 | 0.953 | 0.947 | 0.950 | 0.948 | 0.983 |

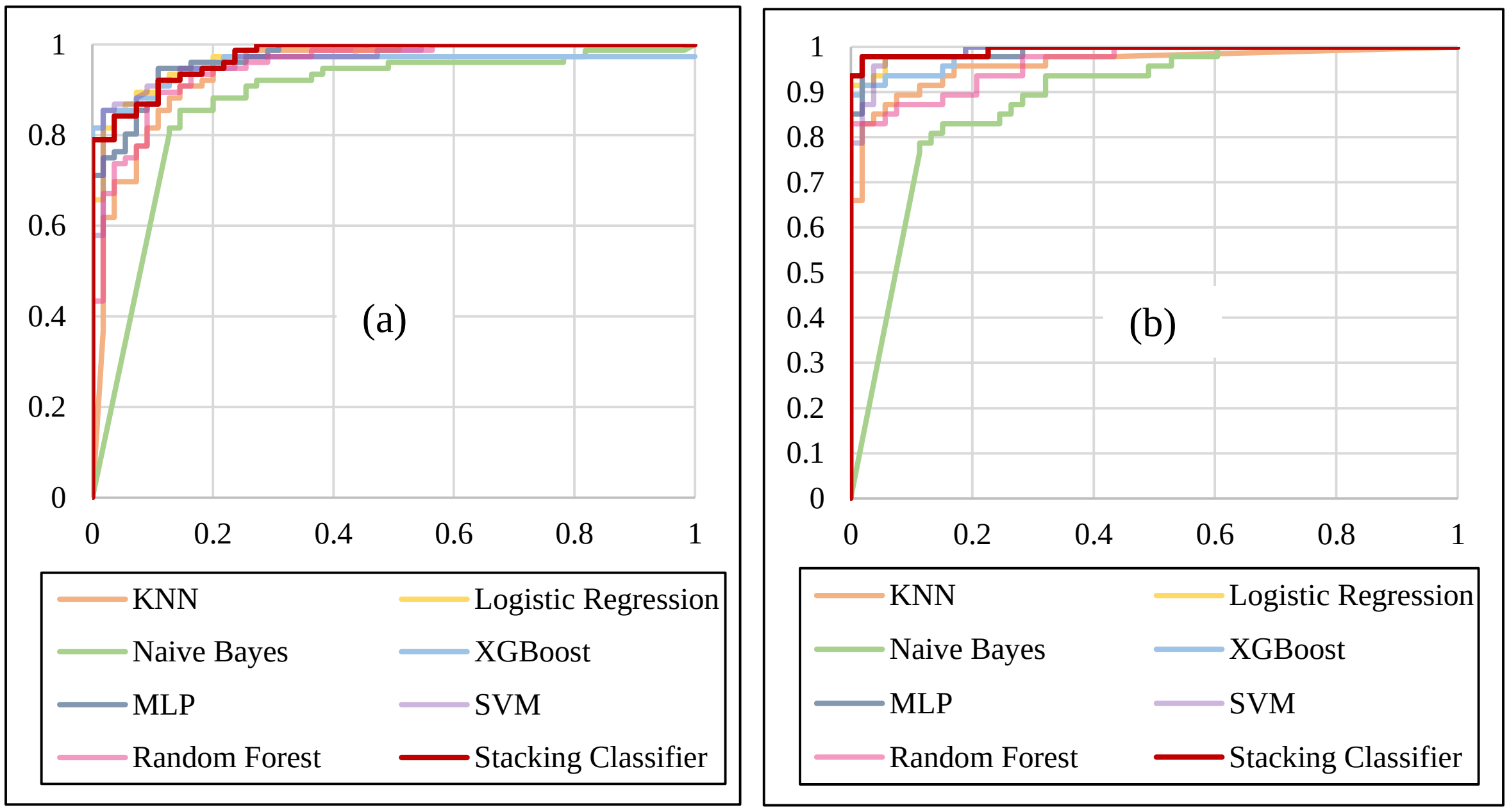

Figure 13: (a) and (b) ROC curve of the BollyBAIT and MVD datasets respectively for the different models used in the stacking classifier.

### 5.2 Feature Importance

We also conducted experiments to gauge the influence of different feature component combinations on the final classification. Table 5 shows the accuracy, f1-score, precision, recall, and ROC-AUC value using our Stacking Classifier on the self-curated BollyBAIT dataset with individual feature components as well as different feature combinations as shown in the table. All the metrics are the average of the individual metrics over the 10-fold stratified cross-validation run of the classifier.

Table 5: Results for Different Feature Combinations

| Feature Group | FT | ST | ACC | PRE | REC | F1 | AUC |
|---|---|---|---|---|---|---|---|
| Individual Feature Sets | | | | | | | |
| Baitiness Analysis (B) | 9.6 | 0.093 | 0.732 | 0.717 | 0.906 | 0.799 | 0.749 |
| Lexical Analysis (L) | 12.1 | 0.094 | 0.718 | 0.714 | 0.872 | 0.784 | 0.763 |
| Sentiment Analysis (S) | 12.5 | 0.095 | 0.653 | 0.664 | 0.864 | 0.734 | 0.666 |
| BERT Features (BR) | 25.5 | 0.250 | 0.900 | 0.898 | 0.937 | 0.916 | 0.964 |
| RESNET Features (RS) | 11.9 | 0.110 | 0.853 | 0.867 | 0.887 | 0.876 | 0.913 |
| Title & Thumbnail Disparity Features (TTD) | 8.2 | 0.095 | 0.640 | 0.691 | 0.698 | 0.693 | 0.641 |
| Title & Content Disparity Features (TCD) | 10.7 | 0.097 | 0.658 | 0.702 | 0.729 | 0.714 | 0.732 |
| Content & Thumbnail Disparity Features (CTD) | 12.6 | 0.097 | 0.587 | 0.592 | 0.950 | 0.729 | 0.493 |
| Combination of Feature Sets | | | | | | | |
| B+L+S | 13.2 | 0.097 | 0.802 | 0.790 | 0.914 | 0.845 | 0.849 |
| BR+RS | 18.7 | 0.119 | 0.899 | 0.909 | 0.922 | 0.915 | 0.959 |
| TTD+TCD+CTD | 14.6 | 0.100 | 0.695 | 0.73 | 0.757 | 0.745 | 0.751 |

We also evaluated feature importance in terms of the number of features selected for classification, and the results obtained are displayed in Table 6. We found that the highest accuracy of 92.89% was achieved when 825 features were chosen using the ANOVA F-value statistical test, the graph for which is shown in Figure 14.

Table 6: Results for Feature Selection

| Selected Features | Accuracy Obtained | Selected Features | Accuracy Obtained |
|---|---|---|---|
| 5 | 79.9934 | 700 | 92.0387 |
| 50 | 88.6620 | 825 | 92.8964 |
| 100 | 90.0372 | 900 | 92.3405 |
| 200 | 90.9650 | 1000 | 91.4312 |
| 500 | 90.6643 | 2000 | 91.2657 |

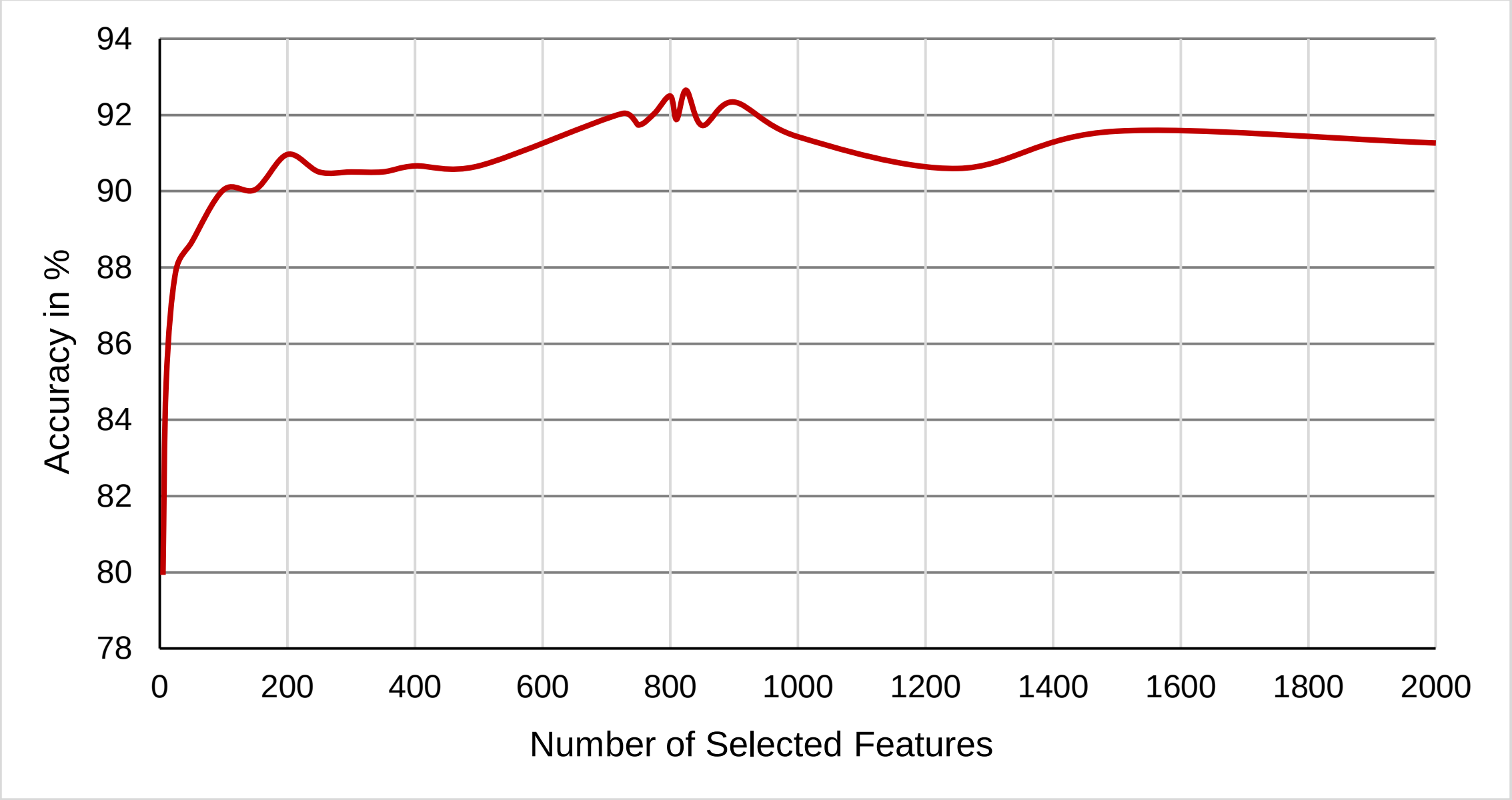

Figure 14: Feature importance graph for the BollyBAIT dataset that peaks at 825 features and 92.89% accuracy

On performing a frequency analysis of all the clickbait titles present in BollyBAIT, we found beguiling words and phrases such as “Official trailer”, “shocking”, “secret”, “Bollywood”, “Shahrukh Khan” that are included in the Lexical analysis of title component, to be prevalent and used frequently, as shown in Figure 15.

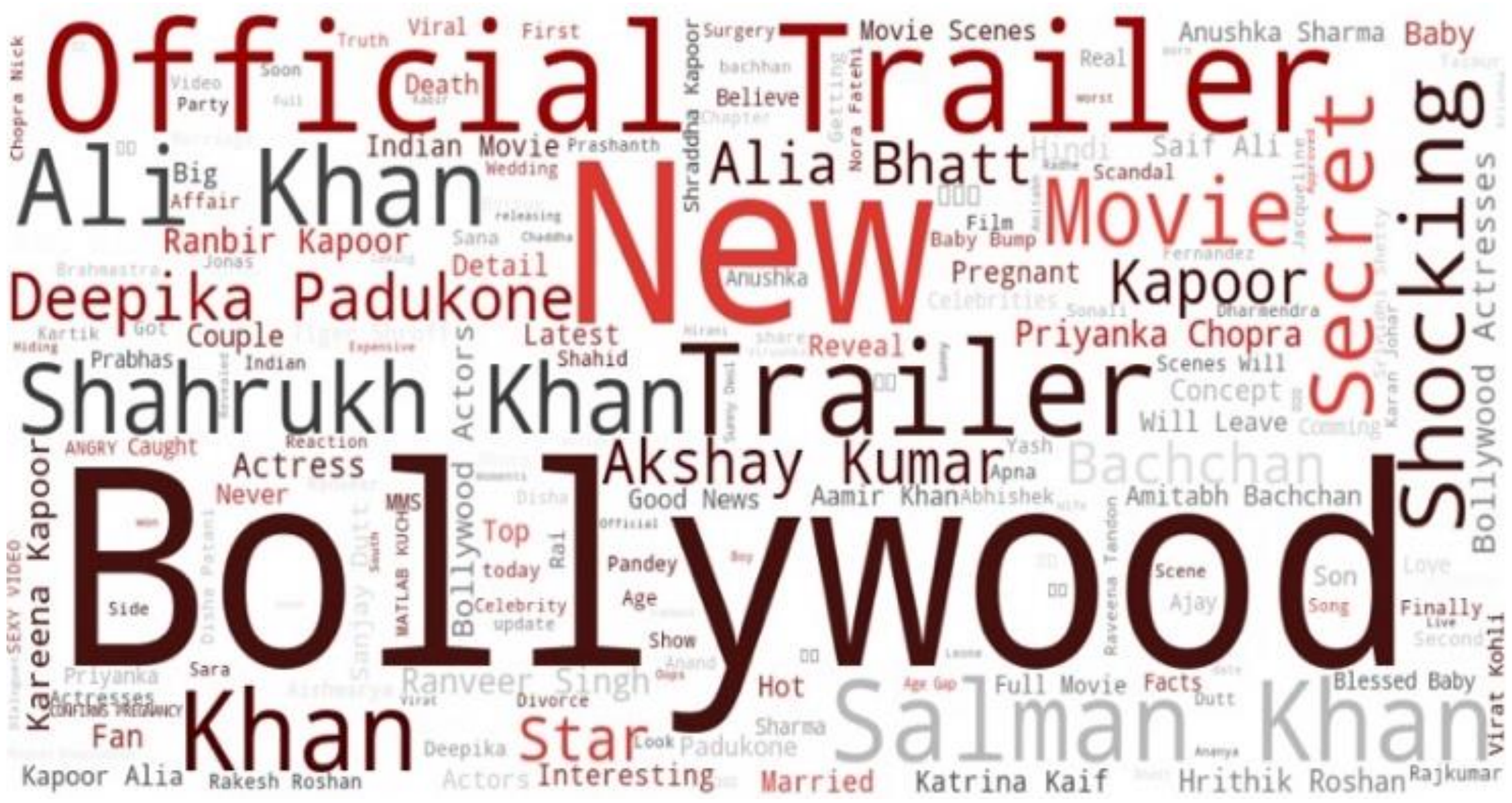

Figure 15: Word cloud of clickbait titles in the BollyBAIT dataset

### 5.3 Clickbait Categories

The BollyBAIT dataset consists of another set of manual labels describing if the video falls under a particular category of clickbait as defined in Section 1. These different types include: ‘Misleading Video’, ‘Spam Content’, ‘Exaggerated Video’, ‘False Promises’ and ‘Exploits Curiosity Gap’. Clickbait classification for each type of clickbait was done based on these categories. Table 7 shows the accuracy, f1-score, precision, recall, and ROC-AUC value for each clickbait type using our stacking classifier. All the metrics are the average of the individual metrics over the 10-fold stratified cross-validation run of the classifier. Figure 16 shows the correlation matrix for the five different clickbait categories.

Table 7: Results for Different Clickbait Categories

| Type of Clickbait | FT | ST | ACC | PRE | REC | F1 | AUC |
|---|---|---|---|---|---|---|---|
| Misleading Video | 33.81 | 0.2041 | 0.8530 | 0.723 | 0.7679 | 0.7412 | 0.9193 |
| False Promises | 36.50 | 0.2061 | 0.8378 | 0.702 | 0.7873 | 0.7384 | 0.8911 |
| Exaggerated Video | 35.65 | 0.2078 | 0.8299 | 0.661 | 0.6897 | 0.6706 | 0.8881 |
| Spam Content | 32.74 | 0.2011 | 0.8682 | 0.781 | 0.8207 | 0.7976 | 0.9352 |
| Exploits Curiosity Gap | 30.55 | 0.2028 | 0.8897 | 0.716 | 0.7109 | 0.7055 | 0.9193 |

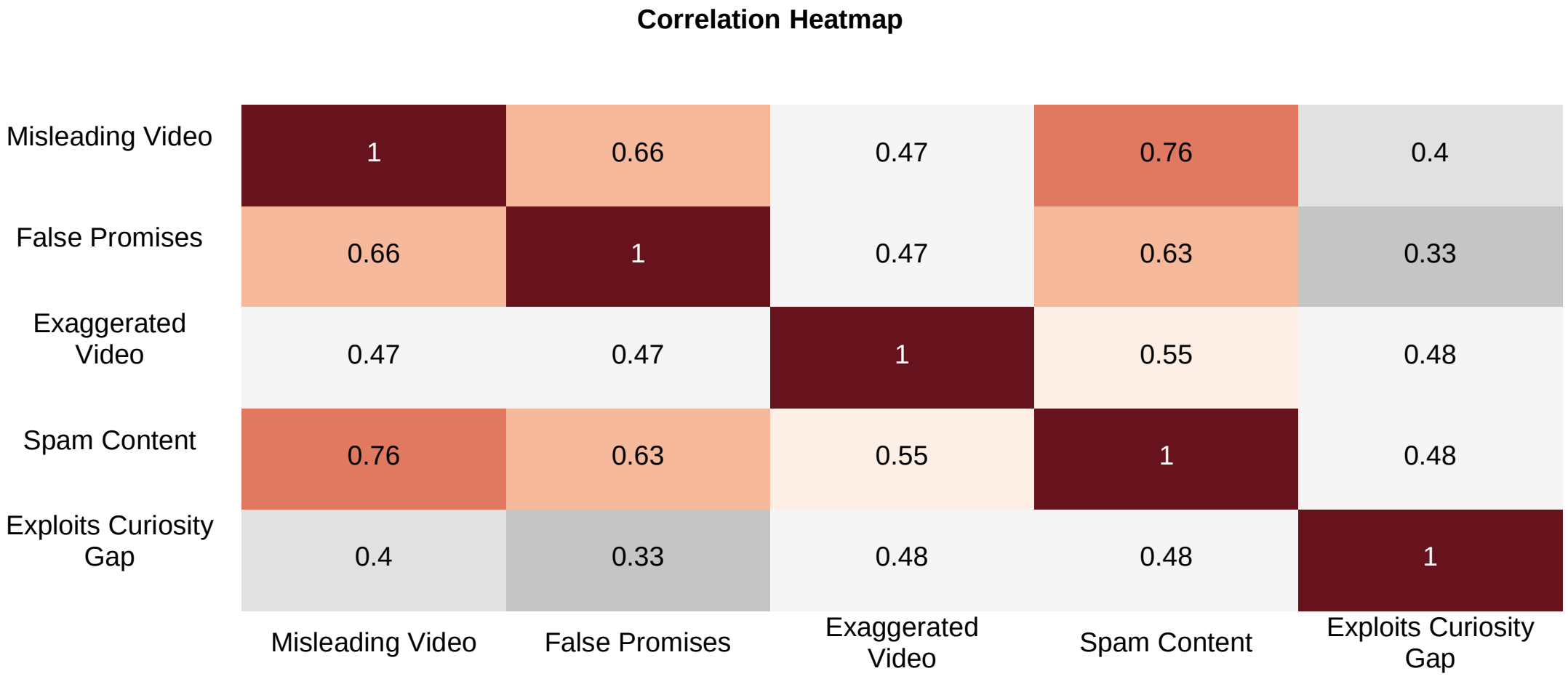

Figure 16: Correlation matrix for the different clickbait categories

## 6 CONCLUSION

In this paper, we present a novel clickbait prevention and detection model that considers the video's actual audio and video content. We also develop a Bollywood-focused BollyBAIT clickbait dataset containing 1000 videos annotated for five different types of clickbait and a categorical clickbait or non-clickbait label. CPDM leverages video descriptors of other modalities to extract video features effectively and comprehensively for clickbait classification:

- We extract the video frames, audio, title, and other data related to the video.
- We obtain features from the teaser image, i.e., the thumbnail, and the teaser text, i.e., the video's title.
- We perform feature mining across image, text, and audio features by applying deep learning techniques and derive feature sets from different components.
- We implement late fusion on the results of feature set mining for binary classification of YouTube videos as clickbait and non-clickbait using a stacking classifier framework.
- We validate the performance of our CPDM framework by testing it across two clickbait-focused datasets, which gave us an accuracy of 0.953 for the MVD dataset and 0.928 for the self-curated BollyBAIT dataset.

The prime advantage of our work is that because the proposed framework is not dependent on meta-features, it is possible to deploy the model on videos from the moment they are posted. This would decrease the number of clickbait videos circulating by filtering them out from the recommendation feed early, thus improving user experience and engagement. Also, the model considers the actual video content and compares it with the teaser text and image, and the essence of clickbait lies in this comparison. Further work can be done by creating large-scale datasets for clickbait classification and extending CPDM to create a real-time plugin to detect clickbait videos across different video-streaming platforms, not just YouTube.